# Experimental determination of phonon thermal conductivity and Lorenz ratio of single crystal bismuth telluride

Mengliang Yao, [1] Stephen Wilson, [2] Mona Zebarjadi, [3] and Cyril Opeil [1]

[1]*Department of Physics, Boston College, Chestnut Hill, Massachusetts 02467, USA*

[2]*Materials Department, University of California, Santa Barbara, California 93106, USA*

[3]*Department of Electrical and Computer Engineering and Department of Materials Science and Engineering, University of Virginia, Charlottesville, Virginia 22904, USA*

We use a magnetothermal resistance method to measure the lattice thermal conductivity of a single crystal of $Bi_2Te_3$ from 5 to 60 K. We apply a large transverse magnetic field to suppress the electronic thermal conduction while measuring thermal conductivity and electrical resistivity. The lattice thermal conductivity is then calculated by extrapolating the thermal conductivity versus electrical conductivity curve to a zero electrical conductivity value. Our results show that the measured phonon thermal conductivity follows the $e^{\Delta_{min}/T}$ temperature dependence and the Lorenz ratio corresponds to the modified Sommerfeld value in the intermediate temperature range. Our low-temperature experimental data and analysis on $Bi_2Te_3$ are an important compliment to previous measurements of Goldsmid [14] and theoretical calculations by Broido *et al.* [21] at higher temperature 100 - 300 K.

## I. Introduction

For most materials at low temperatures their thermal transport $\kappa_{tot}$ involves two types of carriers: one ($\kappa_e$) is the electron (or hole) contribution which conducts both charge and heat, and the other ($\kappa_{ph}$) is the phonon (or equivalently lattice) contribution which only conducts heat; normally it is not easy to separate $\kappa_e$ or $\kappa_{ph}$ from $\kappa_{tot}$. [1-3] In our previous work, we successfully applied the magneto-thermal resistance (MTR) method to three different single crystal metals (Al, Cu, and Zn) by suppressing $\kappa_e$ in a strong magnetic field and extracted $\kappa_{ph}$ from the extrapolation. [4] Unlike metals, which reveal an electron dominant thermal transport ($\kappa_e \gg \kappa_{ph}$), semiconductors and semimetals are completely different where these two components are comparable ($\kappa_{ph} \sim \kappa_e$) or even phonon dominant ($\kappa_{ph} > \kappa_e$). [3, 6] As a result, it is germane to verify the MTR method on materials over than metals, such as single crystal semiconductors, in the same temperature range, of which the thermal transport are usually phonon dominant. Today, many thermoelectric (TE) materials utilize a nanocomposite structure and are nearly impossible to model using fully first principles calculation approaches because of their nearly random matrix structure. Here we explore a single crystal of $Bi_2Te_3$ in order to encourage theoretical modeling of this and other TE materials. Using our method, we can extract lattice thermal conductivity as opposed to total thermal conductivity, and our results could be directly compared to first principle calculations.[4]

When applying the MTR method, there are several important quantities for describing the transport process. The deflecting angle $\gamma$ is defined as the deviation of an electron away from its previous linear motion under the influence of an applied magnetic field between two consecutive collisions and is used to identify the strength of the field [5]

$$\gamma \equiv \omega_c \tau = \mu B = \frac{\sigma_0 B}{ne}$$

(1)

where $\omega_c$ is the cyclotron frequency, $\tau$ is the relaxation time, $\mu$ is the mobility, $B$ is the magnetic induction strength, $\sigma_0$ is the zero field electrical conductivity, $n$ is the carrier concentration, and $e$ is the elementary charge. When $\gamma = \mu B \sim 1$, a significant suppression of the electronic conduction can be observed. [3, 5] Here we define a threshold field $B_{th} = 1/\mu$ as the inverse of the mobility to represent the field one needs to see the suppression, which depends on the temperature and this threshold field decreases when temperature is reduced. In our previous paper, we found that when $T < 50$ K the threshold field $B_{th}$ becomes smaller than our maximum applied field (10 T) for metals. As semiconductors usually have larger mobilities than metals, it is easier to observe the suppression of electronic conduction for semiconductors, *e.g.* our $Bi_2Te_3$ single crystal. [4]

In zero magnetic field the Lorenz ratio is defined as,

$$L \equiv \frac{\kappa_e}{\sigma_0 T}$$

(2)

where $\kappa_e$ is the electrical thermal conductivity, $\sigma_0$ is the electrical conductivity, and $T$ is the absolute temperature, all the quantities are measured in zero field. In practice the Lorenz ratio is used to estimate the electronic contribution to the total thermal conductivity from the electrical resistivity measurements. [7-10] The standard value of the Lorenz ratio is the Sommerfeld value $L_0 = \frac{\pi^2}{3}\left(\frac{k_B}{e}\right)^2 = 2.443 \times 10^{-8}$ V$^2$/K$^2$. This is the Wiedemann-Franz law and it is valid for metals at high temperatures and when neglecting the thermoelectric term. [1] However, for semiconductors or typical TE materials such as $Bi_2Te_3$, or $Bi_2Se_3$ where the Seebeck coefficients are on the order of 100 µV/K at room temperature, the relation between the Lorenz ratio and its Sommerfeld value needs to be modified to include the thermoelectric term as [1]

$$L + S^2 = L_0$$

(3)

Although several results have been reported in the literature using magnetic field to suppress electronic conduction, [6, 11] the applied fields were insufficient to achieve the threshold field $B_{th}$ throughout the experimental temperature range. It is possible to choose both a temperature range and applied magnetic field for a particular material that allows for complete suppression. For our experiments, we use a Physical Properties Measurement System (PPMS) from Quantum Design and are able to control the magnetic field and the temperature simultaneously so that the applied magnetic field is sufficient to produce the necessary threshold field $B_{th}$ throughout the entire experimental temperature range (5 ~ 60 K). Under these conditions we are able to indirectly measure the phonon thermal conductivity with a virtual suppression of electronic thermal conductivity.

## II. Experimental Details

Our samples are taken from a large *p*-type stoichiometric single crystal $Bi_2Te_3$ (001) grown by the Bridgman method and was determined to have a residual resistance ratio (RRR, defined by $\rho_{300K}/\rho_{2K}$) of ~30. The crystal structure (001) was confirmed by X-ray diffraction using a Bruker D2 PHASER. Specimens are cut into typical dimensions of $1 \times 3.5 \times 10$ mm$^3$. Because silver can migrate into bismuth compounds, silver paint or epoxy were not used to make electrical/thermal contacts to the sample. [17-19] Due to the laminar structure of Bi$_2$Te$_3$ and the thickness of our specimen, soldering to sputtered metallic contacts on the surface of our samples was not considered sufficient for thermal contacts in our measurements. As a result we built four miniature brass compression clamps with 1 mm screws to hold the sample for thermal transport measurements ($\kappa_{tot}$ and $S$). To ensure thermal contact between the brass clamp and sample indium dots were pressed between the clamp and the sample. Gold coated OFHC copper flat wires, supplied by Quantum Design, were soldered onto the brass clamps and then attached to the TTO puck. Electrical contacts for resistivity and Hall measurements ($\rho_{xx}$ and $\rho_{yx}$) were made by pressing indium dots onto the sample. A standard 4-probe method was applied for all of electrical conductivity measurements.

Thermal conductivity and Seebeck coefficient measurements were performed using the thermal transport option (TTO) of the Quantum Design PPMS. The sample is placed in a transverse orientation where the heat flow is perpendicular to the magnetic field, which was oriented along the c-axis of Bi$_2$Te$_3$ single crystal. When performing the experiments, we first set the magnetic field and scan the temperature from 60 K to 5 K at a rate of 0.5 K/min. The field is then set to a different fixed value and the conductivity measurement is repeated while increasing the temperature from 5 - 60 K. Thermal conductivity data was taken across the temperature region in integer values of magnetic field from 0 T to 9 T. A similar sequence is used for the electrical resistivity and Hall measurements using an LR-700 AC resistance bridge from Linear Research Inc., where the temperature was fixed and the magnetic field was scanned. For these measurements, our sample was mounted with the same orientation to the field used in the thermal transport measurements. Propagated measurement errors are calculated from the standard deviation and determined to be 3% for $\kappa$, 6% for $S$, 0.5% for $\rho$, and derived to be 0.7% for $\sigma$, 3% for $\kappa_{ph}$, and 18% for the Lorenz ratio.

### III. Magneto-transport Measurements and Analysis

Resistance in zero magnetic field can be considered as a scalar quantity and the reciprocal of electrical resistivity $\rho_{xx}$ is electrical conductivity $\sigma_{xx}$. However, when applying magnetic field to a material and measuring its resistance, the transport coefficients are no longer scalars but considered as tensor quantities, as such $\sigma_{xx} \neq 1/\rho_{xx}$. Similar to our previous work of magneto-resistance measurements on single crystal metals [4], the transport coefficients of Bi$_2$Te$_3$ need to be considered as tensor quantities. While it is difficult to measure the Hall resistance of single crystal metals because of their low resistivity, it is possible to measure the Hall resistance of semiconductors such as Bi$_2$Te$_3$. Hall measurements can be performed here due to their particular combination of carrier concentration and mobility. Therefore, by measuring resistivity under field and the electrical conductivity component $\sigma_{xx}$ can be calculated by using the following formula: [5, 12, 13]

$$\sigma_{xx} = \frac{\rho_{xx}}{\rho_{xx}^2 + \rho_{yx}^2}$$

(4)

where $\rho_{xx}$ and $\rho_{yx}$ are the normal electrical resistivity and the Hall resistivity, respectively. In order to extract phonon thermal conductivity $\kappa_{ph}$ from the total thermal conductivity $\kappa_{tot}$, one needs to plot $\kappa_{xx}$ vs. $\sigma_{xx}$ and extrapolate to the intercept to obtain $\kappa_{ph}$. When extrapolating from the $\kappa_{xx} \sim \sigma_{xx}$ curve to determine $\kappa_{ph}$, the following formula is used [4]

$$\kappa_{tot}(\sigma) = \kappa_{ph} + \frac{\kappa_0 - \kappa_{ph}}{1 + \lambda^2 \left(\frac{\sigma_0}{\sigma} - 1\right)}$$

(5)

where $\kappa_{tot}$ and $\sigma$ are the measured values of total thermal conductivity and the derived values of electrical conductivity according to eqn. (4); $\kappa_0$ and $\sigma_0$ are the above values in zero field; $\kappa_{ph}$ is one of the two fitting parameters representing the phonon thermal conductivity, while $\lambda$ is the other fitting parameter representing the ratio between the thermal and drift mobilities. If $\lambda = 1$ then $\kappa_{tot}$ and $\sigma$ are linearly related, which can be observed at high temperature,[3] however, in general this condition is not valid.

Fig. 1 shows the total thermal conductivity and Seebeck coefficient measurements in zero field, as well as the extracted carrier concentration from 5 K through 300 K. In Fig. 1(a) we also include the results from Goldsmid's experiments [14] and Broido et al.'s first principle calculations [21] on $Bi_2Te_3$ through the temperature of 100-300 K. The data points show they coincide with each other very well, and our data below 100 K cover the range that is not covered and show that our data tend to theirs. In Fig. 1(b) the Seebeck curve shows a linear relationship with temperature below 250 K, while the calculated carrier concentration maintains nearly constant above 200 K.

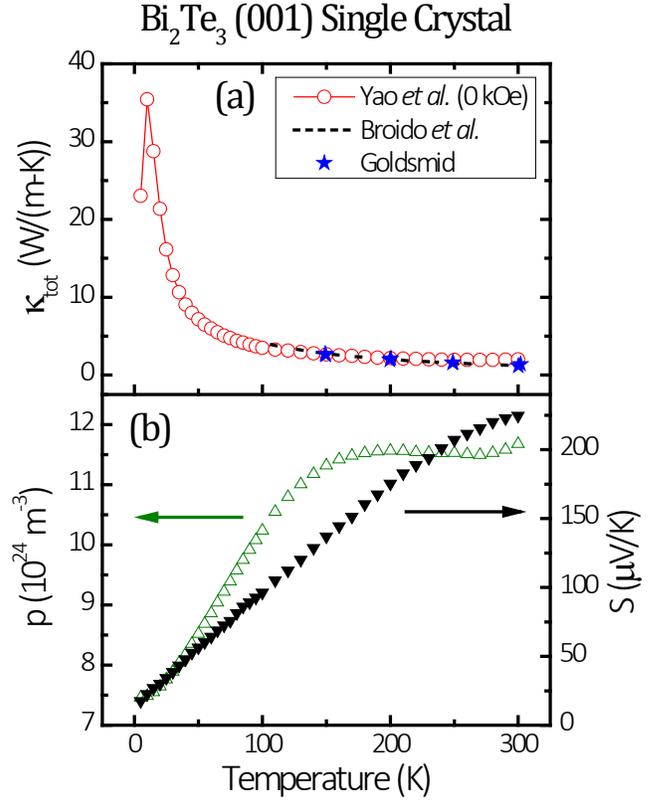

FIG. 1: 1(a) Here we compare our thermal conductivity data (red open circles), Goldsmid's experimental data (blue stars, from ref. 14) and the calculation of phonon thermal conductivity from Broido *et al.* (black dashed line, from ref. 21). 1(b) Our Seebeck coefficient data (black downward triangles) are shown with the carrier concentration (green upward triangles) extracted for 5 - 300 K. The carrier concentration is calculated from the Hall measurements following the Drude model.

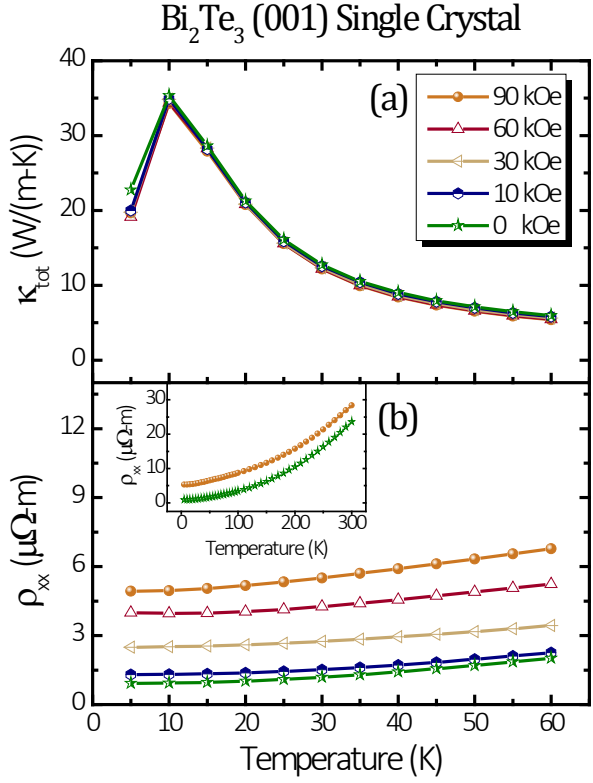

FIG. 2: Thermal conductivity 2(a) and electrical resistivity 2(b) are plotted against temperature between 5 K and 60 K in different magnetic fields. The error bars for thermal conductivity data are ~3% and resistivity error bars are ~0.5% both appear smaller than the data symbols. The inset for 2(b) illustrates the magnetoresistance of $Bi_2Te_3$ up to room temperature where the difference in magnetoresistance remains approximately constant in 0 T and 9 T.

Fig. 2 shows the temperature dependent behavior of the total thermal conductivity and electrical resistivity under various applied magnetic fields. Due to the low carrier concentration as compared with metals and the high mobility of holes, the effects of the magnetoresistance are easily observed even at room temperature, see Fig 2(b) inset. On the contrary, the thermal conductivity is not significantly suppressed even when the threshold field $B_{th}$ is much smaller than the applied magnetic field. The cause of this behavior lies in phonon dominance of the thermal transport of the single crystal $Bi_2Te_3$. When comparing TE nanocomposite materials with the single crystals, the latter have fewer impurities and defects and as a result only a small portion of the phonons are scattered during the transport process.

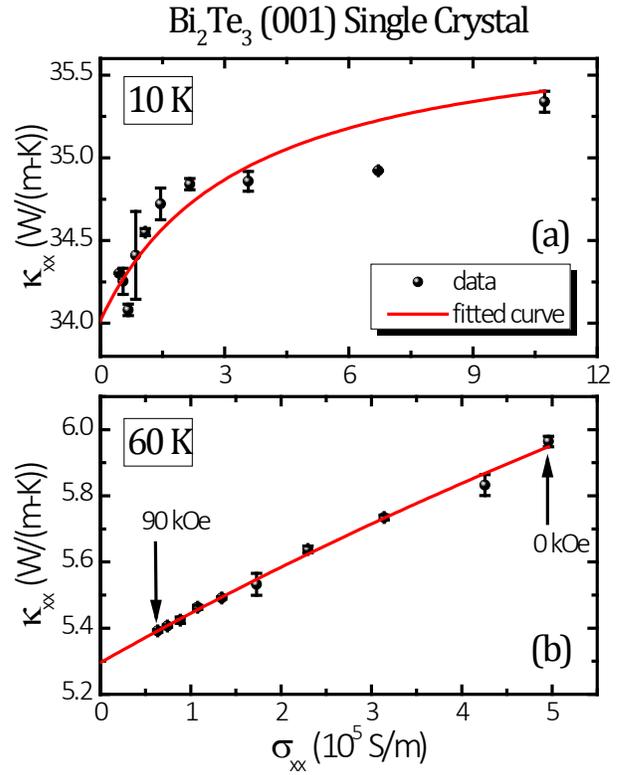

FIG. 3: Total thermal conductivity is plotted against electrical conductivity at 10 K and 60 K. The red fitted curve is calculated according to eqn. (5).

In order to extract $\kappa_{ph}$ from $\kappa_{tot}$, one needs to plot the measured thermal conductivity vs. the calculated electrical conductivity curve for a specific temperature. Fig. 3 shows a set of $\kappa_{tot} \sim \sigma$ curves at 10 K and 60 K from which we can extrapolate the $\kappa_{ph}$ from $\kappa_{tot}$. As shown in the literature [3] $\kappa_{xx}$ and $\sigma_{xx}$ have a linear relationship at higher temperatures because the ratio of thermal mobility $\mu_t$ and drift mobility $\mu_d$ approaches unity. This linearity is verified by our 60 K Bi$_2$Te$_3$ data shown in Fig. 3(b). We note that the 10 K data in Fig. 3(a) is non-linear but eqn. (5) can be used to fit the data over the range of electrical conductivity. The intercept from the extrapolation to zero electrical conductivity corresponds to the pure phonon thermal conductivity $\kappa_{ph}$, which is assumed to be field independent and is a condition of our analysis. In this sense we able to separate the electronic and phonon thermal transport by applying a sufficiently high magnetic field across our experimental temperature range.

In Fig. 4(a) the phonon thermal conductivity $\kappa_{ph}$ and total thermal conductivity at zero field $\kappa_{tot}(0\ \text{kOe})$ are plotted together. The upper right inset shows the proportion of phonon thermal conductivity to the total thermal conductivity, as we see for most temperature points the ratio is above 90% (indicating that the specimen is of high quality with few impurities and defects), only at the lowest temperature the ratio goes down. This trend is also easily understood from the behavior of heat capacity of phonons and electrons, since at low temperatures $\kappa_{ph} \sim T^3$ while $\kappa_e \sim T$. Therefore as a result $\kappa_{ph}$ drops more quickly than $\kappa_e$ in the low temperature range and its ratio to the total thermal conductivity decreases.

The dimensionless Lorenz ratio, defined as $L/L_0$, is shown in Fig. 4(b). The smooth dark cyan curve models the modified Sommerfeld value according to eqn. (3) considering our Seebeck coefficient and thermoelectric effect data. The thickness of the smooth cyan curve indicates the error boundary spread due to the uncertainty from our Seebeck coefficient measurements. The upper right inset to Fig. 4(b) shows a more detailed data comparison between 20 K and 60 K within this temperature range and the experiment demonstrates a high correlation between our data and the modified Wiedemann-Franz law. The only inconsistency happens at 10 K, with its value going beyond unity (see further discussion in Sect. IV).

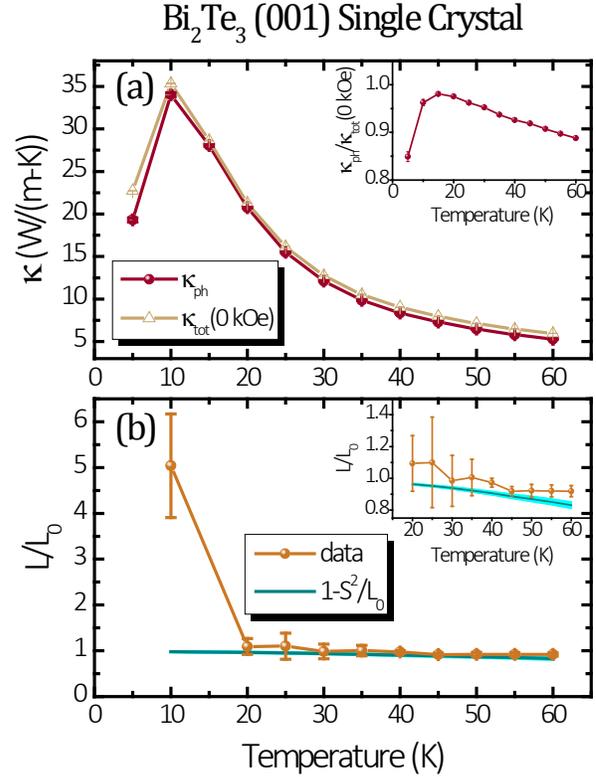

**FIG. 4: Phonon thermal conductivity 4(a) and dimensionless Lorenz ratio 4(b) are plotted against temperature. $1 - S^2/L_0$ is the modified Sommerfeld value according to eqn. (3). The inset to 4(a) shows the ratio $\kappa_{ph}/\kappa_{tot}(0\ \text{kOe})$ across the same temperature range. The inset to 4(b) highlights the calculations from our data $L/L_0$ and the modified W-F law from $20\ \text{K} \leq T \leq 60\ \text{K}$.**

## IV. Discussion

In Fig. 4 we see in the temperature range 20 - 60 K the Lorenz ratio is very close to its standard Sommerfeld value if the thermoelectric effect is considered, suggesting for our $Bi_2Te_3$ single crystal the Wiedemann-Franz law is satisfied. Thus across this temperature range, the relaxation time of electrons in the thermal process deviates little from the relaxation time in the electrical process. This results from scattering and depends on the impurity concentration expressed in the measured RRR of ~30 for our $Bi_2Te_3$ single crystal. We note that for single crystal metals with a RRR one order of magnitude higher than found for this semiconductor, the Lorenz ratio can differ significantly as shown in our previous paper. [4]

However, at 10 K our calculated Lorenz ratio deviates significantly from the expected Sommerfeld value. One explanation for this sudden increase in the Lorenz ratio is the laminar structure of $Bi_2Te_3$ single crystal. Normally due to the scattering process, the mean free path of the electrons is smaller than the layer distance of $Bi_2Te_3$ sheets; as a result the specimen behaves as a 3-dimensional bulk material. However, as $T \to 0$ K, the mean free path of electrons becomes greater than the layer distance, while the electrons are restricted to the 2-dimensional structure. This regime transformation may result in a different Lorenz ratio from the Wiedemann-Franz law. Consistent with our results a notable violation of the Wiedemann-Franz law has been observed by Wakeham et al. [20] for $Li_{0.9}Mo_6O_{17}$ for $T < 20$ K in the one dimensional case. However, more low-temperature investigations on quasi 1-D and 2-D materials are needed to confirm these findings. Finally, in the literature it is reported that $Bi_2Te_3$ single crystals may have a dimensionless Lorenz ratio much larger than unity, [14-16] however, those abnormalities are due to the bipolar contributions and the emerging temperature is quite different from and much higher than ours. We note the carrier concentration in Fig. 1(b) increases steadily with rising temperature suggesting that there is no bipolar contribution in our experimental regime (5 - 60 K). Using the estimation of effective mass given by Harman et al., [22] we calculate the Fermi level to be 95 – 125 meV inside the valence band throughout the whole temperature range (5 – 300 K).

In order to extrapolate $\kappa_{ph}$ and utilize eqn. (5) with a high degree of precision, the data points must be in close proximity to the intercept, see Fig. 3. Fortunately for our $Bi_2Te_3$ samples the threshold field is already achieved even at room temperature and as a result the 90 kOe data point is close to the intercept for both the 10 K and 60 K figures. The proximity of these data points to the intercept guarantees the reliability of our extrapolation.

The temperature behavior of $\kappa_{ph}$ depends on the underlying scattering mechanism. In our $Bi_2Te_3$ single crystals the thermal transport is dominated by the phonons and we can consider three distinct temperature regions that effect scattering. In the first temperature region one considers the phonon-phonon scattering of the Umklapp process at high temperatures ($T \gg \Theta_D$) where we note a $T^{-1}$ behavior. In the second region or intermediate ($T \sim \Theta_D$) temperature range we have a $e^{\Delta_{min}/T}$ dependence for the same phonon-phonon scattering. [1] And finally we have a $T^3$ dependence due to the heat capacity of phonons at low temperatures ($T \ll \Theta_D$) and this results in a peak in the $\kappa_{ph} \sim T$ curve at ~10 K for our $Bi_2Te_3$ single crystal.

Fig. 5 shows the $\kappa_{ph} \sim T$ curve in a *log-log* scale, we note that at lowest temperatures the curve deviates from the $T^3$ power law behavior. This deviation results from the temperature limit of our data to 5 K. If it were possible to measure $\kappa_{tot}$ from 1 - 5 K then likely the $T^3$ law would be verified. (The effective limit of thermal conductivity measurements with the QD-PPMS is 5 K.) After all, compared with our data on single crystal metals,[4] the peak temperature of $\kappa_{ph}$ of Bi$_2$Te$_3$ is much lower. In the higher temperature regime $\kappa_{ph}$ begins to follow the $e^{\Delta_{min}/T}$ temperature dependence, as shown in Fig. 5.

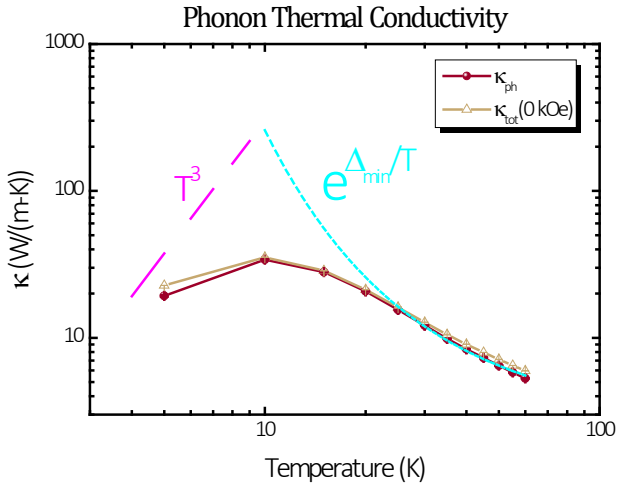

**FIG. 5: Phonon thermal conductivity of Bi$_2$Te$_3$ single crystal is plotted against temperature with a *log-log* scale. The two dashed lines are guides to the eye showing different temperature dependences.**

The size of the threshold field $B_{th}$ takes on a key role in the MTR method when attempting to suppress electronic thermal conductivity behavior. If the applied field is smaller than the threshold field over a particular temperature range, then the suppression is not sufficient to extrapolate a $\kappa_{ph}$ value. To achieve suppression, a specimen of high mobility is required, which implies that the RRR of the sample cannot be small. In a parallel study we used the MTR method on a Bi$_2$Se$_3$ single crystal with a RRR of only ~2 compared to RRR of ~30 for Bi$_2$Te$_3$. The Bi$_2$Se$_3$ single crystal has a very low mobility compared to the Bi$_2$Te$_3$ (~10 %), and thus the threshold field is never reached even at the lowest temperature. As a result we determined that this Bi$_2$Se$_3$ sample with a RRR ~2 was not a viable candidate to apply the MTR method.

Generally, semiconductors exhibit larger Hall resistance than found in pure metals and their electrical conductivity can be calculated by eqn. (4). We note that the electrical resistivity $\rho_{xx}$ always increases, see Fig. 2, when temperature goes up no matter what the magnetic field applied, however, the electrical conductivity $\sigma_{xx}$ in Fig. 6 does not always decrease. Normally, we would expect $\sigma_{xx}$ to decrease with temperature while $\rho_{xx}$ increases with temperature. However, in Fig. 6 we observe a crossover behavior in the $\sigma_{xx} \sim B$ curves at about 27.5 kOe such that as temperature increases both $\sigma_{xx}$ and $\rho_{xx}$ increase in a uniform fashion. This results from a large Hall resistance in our Bi$_2$Te$_3$ sample and indicates that the cyclotron frequency, $\omega_c$, of the Landau levels due to the field becomes larger than the collision frequency $1/\tau$ as the crossover emerges.

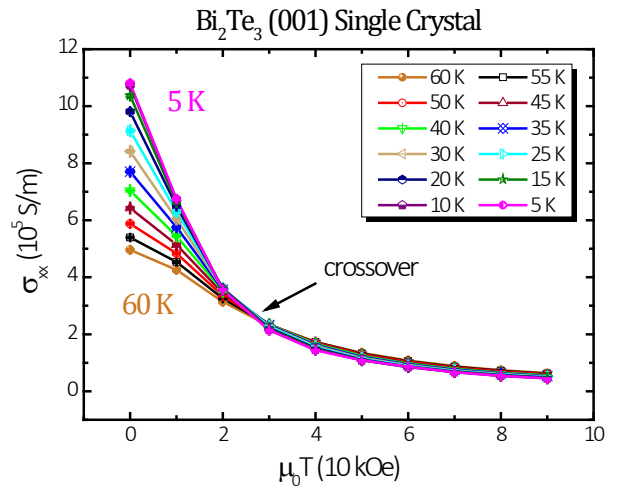

**FIG. 6: Transverse electrical conductivity is plotted against applied magnetic field at different temperatures. A crossover is observed around 27.5 kOe.**

## V. Conclusions

In this study, the phonon thermal conductivity and Lorenz ratio of a $Bi_2Te_3$ single crystal are extracted from the total thermal conductivity using the magnetothermal resistance (MTR) method. We demonstrate experimentally that the phonon thermal conductivity for $Bi_2Te_3$ follows the $e^{\Delta_{min}/T}$ temperature dependence over the intermediate temperature range. We note the sudden rise in the $L/L_0$ ratio at 10 K (see Fig. 4) seems to violate the W-F law. We propose that this apparent violation of the W-F law is related to the reduced dimensionality considerations of this material as $t \to 0$ K. The Lorenz ratios at other temperatures obey the modified Wiedemann-Franz law very well. These results suggest that a comparison with a first principle calculation in this temperature range may be helpful to understand the physics of this and other similar thermoelectric materials.

## Acknowledgements

This work was supported by Solid State Solar - Thermal Energy Conversion Center (S3TEC), an Energy Frontier Research Center funded by the U. S. Department of Energy, Office of Science, Office of Basic Energy Science under award number DE-SC0001299. C.P.O. would like to thank Robert D. Farrell, S.J. and Christopher Noyes for helpful comments on the manuscript, and acknowledges financial support from the Trustees of Boston College. M.Y. is grateful to Ying Ran for helpful discussions and comments on the manuscript. The work at the University of Virginia is supported by the Air Force Young Investigator Award, grant number FA9550-14-1-0316. Stephen Wilson acknowledges support from NSF CAREER Grant No. DMR-1056625.

## References

[1] E. M. Lifshitz and L. P. Pitaevskii, *Physical Kinetics* (Butterworth-Heinemann, Oxford, 1981).

[2] Terry M. Tritt, *Thermal Conductivity: Theory, Properties, and Applications* (Kluwer Academic / Plenum Publishers, New York, 2004).

[3] K. C. Lukas, W. S. Liu, G. Joshi, M. Zebarjadi, M. S. Dresselhaus, Z. F. Ren, G. Chen, and C. P. Opeil, Phys. Rev. B **85**, 205410 (2012).

[4] M. Yao, M. Zebarjadi, and C. Opeil, unpublished.

[5] A. B. Pippard, *Magnetoresistance in Metals* (Cambridge University Press, Cambridge, 1989).

[6] C. Uher, and H. J. Goldsmid, Phys. Stat. Solidi (b) **65**, 765 (1974).

[7] Qian Zhang, Yucheng Lan, Silong Yang, Feng Cao, Mengliang Yao, Cyril Opeil, David Broido, Gang Chen, and Zhifeng Ren, Nano Energy **2**, 1121 (2013).

[8] Weishu Liu, Chuan Fei Guo, Mengliang Yao, Yucheng Lan, Hao Zhang, Qian Zhang, Shuo Chen, Cyril P. Opeil, and Zhifeng Ren, Nano Energy **4**, 113 (2014).

[9] Weishu Liu, Hee Seok Kim, Shuo Chen, Qing Jie, Bing Lv, Mengliang Yao, Zhensong Ren, Cyril P. Opeil, Stephen Wilson, Ching-Wu Chu, and Zhifeng Ren, Proc. Natl. Acad. Sci. USA **112**, 11, 3269 (2015).

[10] Qian Zhang, Eyob Kebede Chere, Kenneth McEnaney, Mengliang Yao, Feng Cao, Yizhou Ni, Shuo Chen, Cyril

Opeil, Gang Chen, and Zhifeng Ren, Adv. Energy Mater. **5**, 1401977 (2015).

[11] D. Armitage, and H. J. Goldsmid, J. Phys. C **2**, 2138 (1969).

[12] Haijun Zhang, Chao-Xing Liu, Xiao-Liang Qi, Xi Dai, Zhong Fang, and Shou-Cheng Zhang, Nature Phys. **5**, 438 (2009).

[13] S. Ishiwata, Y. Shiomi, J. S. Lee, M. S. Bahramy, T. Suzuki, M. Uchida, R. Arita, Y. Taguchi, and Y. Tokura, Nature Mater. **12**, 512 (2013).

[14] H. Goldsmid, Proc. of the Phys. Soc. Sec. B **69**, 203 (1956).

[15] P. J. Price, Proc. of the Phys. Soc. Sec. B **69**(8), 851 (1956).

[16] N. F. Hinsche, I. Mertig, and P. Zahn, J. of Elec. Mater. **42**, 7, 1406 (2013).

[17] J. D. Keys, and H. M. Dutton, J. Phys. Chem. Solids **24**, 563 (1963).

[18] H. P. Dibbs, and J. R. Tremblay, J. Appl. Phys. **39**, 2976 (1968).

[19] W. Lin, D. Wesolowski, and C. Lee, J. Mater. Sci.: Mater. Electron **22**, 1313 (2011).

[20] N. Wakeham, A. Bangura, X. Xu, J. Mercure, M. Greenblatt, and N. Hussey, Nat. Comm. **2**, 396 (2011).

[21] O. Hellman, and D. Broido, Phys. Rev. B **90**, 134309 (2014).

[22] T. C. Harman, B. Paris, S. E. Miller, and H. L. Georing, J. Phys. Chem. Solids **2**, 181 (1957).